\documentclass[12pt,preprint]{aastex}

\slugcomment{Not to appear in Nonlearned J., 45.}

\shorttitle{Parker instability with cosmic-ray diffusion}
\shortauthors{Kuwabara et al.}

\begin{document}


\title{NONLINEAR PARKER INSTABILITY WITH THE EFFECT OF COSMIC-RAY DIFFUSION}


\author{T. Kuwabara}
\affil{Graduate Institute of Astronomy, National Central University,
       No. 38, Wu-chuan Li, Chung-li, Tao-yuan, Taiwan 320, R.O.C.}
\author{K. Nakamura}
\affil{Matsue National College of Technology, Nishi-ikoma-chou, Matsue-city,
       Shimane 690-8518, Japan}
\and
\author{C. M. Ko}
\affil{Department of Physics, Institute of Astronomy and Center for
       Complex Systems, National Central University,
       Chung-Li, Tao-yuan, Taiwan 320, R.O.C.}



\begin{abstract}
We present the results of linear analysis and two-dimensional local
magnetohydrodynamic (MHD) simulations of the Parker instability,
including the effects of cosmic rays (CRs), in magnetized gas disks
(galactic disks).
As an unperturbed state for both the linear analysis and the MHD simulations,
we adopted an equilibrium model of a magnetized
two-temperature layered disk with constant gravitational
acceleration parallel to the normal of the disk.
The disk comprises a thermal gas, cosmic rays and a magnetic field
perpendicular to the gravitational accelerartion. Cosmic ray diffusion along
the magnetic field is considered; cross field-line diffusion is supposed
to be small and is ignored.
We investigated two cases in our simulations.
In the mechanical perturbation case we add a velocity perturbation
parallel to the magnetic field lines, while in the explosional
perturbation case we add cosmic ray energy into a sphere where the cosmic rays
are injected. Linear analysis shows that the growth rate of the
Parker instability becomes smaller if the coupling between the CR
and the gas is stronger (i.e., the CR diffusion coefficient is smaller).
Our MHD simulations of the mechanical perturbation confirm this result.
We show that the falling matter is impeded by the CR pressure gradient,
this causes a decrease in the growth rate. In the explosional
perturbation case, the growth of the magnetic loop is faster when the coupling
is stronger in the early stage. However, in the later stage the
behavior of the growth rate becomes similar to the mechanical perturbation
case.

\end{abstract}




\section{INTRODUCTION}
It has been suggested that magnetic fields may play important roles for
active phenomena in space, for example in
astrophysical jets \citep[e.g.,][]{shibata85,matsumoto96},
solar activities \citep[e.g.,][]{priest82,yokoyama01}, and active galaxies
\citep[e.g.,][]{kuwabara00}.
With such active phenomena, if a gas layer is supported by the horizontal
magnetic fields against gravity, then the Parker instability may appear and
can play an important role.
Magnetic fields are also thought to play an important role in accretion disks
\citep[e.g.,][]{stella84,kato86} and galactic disks.
For example, in galactic disks, the interstellar matter (ISM)
is aggregated and grows to giant cloud complexes in spiral arms of galaxies
via the Parker instability
\citep[e.g.,][]{mouschovias,mouschovias74,elmegreen82a,elmegreen82b,
elmegreen86}.
On the other hand,
cosmic rays (CRs) may also play an essential role in
the dynamics of the ISM, since it is recognized that the energy density
of cosmic rays is of the same order as that of the magnetic field and
turbulent gas motions \citep{parker69,ginzburg76,ferriere01}.
The importance of the effects of cosmic rays has been acknowledged, and
a discussion concerning CRs was also given in the original work on the
Parker instability \citep{parker66}.
For studying the dynamics of CRs, there are several approaches.
The particle-particle approach treats the plasma and the
CRs as particles; the fluid-particle approach treats the plasma as a fluid and
the CRs as particles; and the fluid-fluid approach treats the plasma and the CRs
as fluids.
The hydrodynamical approach to the Parker instability and Parker-Jeans
instability without CRs has been done by nonlinear calculation
\citep[e.g.,][]{matsumoto88,shibata89,chou00,kim02}.
On the other hand, in spite of the suggestions of many astrophysical
applications, there are very few papers on the evolution of the Parker
instability with the effects of cosmic rays \citep{hanasz97,hanasz00}.
\citet{hanasz00} carried out calculations on the Parker instability
induced by cosmic ray injection from a supernova under the thin-flux-tube
approximation, and suggested that the instability grows on shorter timescales,
with the values of the diffusion coefficient decreasing.
As the diffusion coefficient decreases, the diffusion speed of the CRs
decreases. Since the region where the CR energy is injected keep it for
a long time, the dynamics is dominated by the interactions near the injection region.

In this paper, we present the results of a linear analysis
and MHD simulation of the Parker instability with the effects
of cosmic rays, starting from an equilibrium two
temperature layered disk. We adopt the hydrodynamic approach for
cosmic ray propagation \citep{drury81,ko92}.
The paper is organized as follows.
In \S\,2,
we present our physical assumptions and the equilibrium model.
Linear stability analysis of the system is given in \S\,3,
and the numerical results are described in \S\,4.
\S\,5 provides a summary and discussion.

\section{MODELS}
\subsection{Assumptions and basic equations}
We investigate the Parker instability with the effect of the CRs
in the galactic disk.
The basic equations are the MHD
equations combined with the CR energy equation,
\begin{equation}
\frac{\partial \rho}{\partial t}+\mbox{\boldmath $\nabla$}\cdot
\left( \rho \mbox{\boldmath $V$}\right)=0,
\label{eq:01}
\end{equation}
\begin{equation}
\frac{\partial}{\partial t}\left(\rho\mbox{\boldmath $V$}\right)+
\mbox{\boldmath $\nabla$}\cdot
\left[ \rho \mbox{\boldmath $V$}\mbox{\boldmath $V$}+
\left(P_{\rm g}+P_{\rm c}+\frac{\mbox{\boldmath $B^2$}}{8\pi}\right)I-
\frac{\mbox{\boldmath $BB$}}{4\pi} \right]-\rho\mbox{\boldmath$g$}
+2\rho\mbox{\boldmath $\Omega$}\times\mbox{\boldmath $V$}=0,
\label{eq:02}
\end{equation}
\begin{equation}
\frac{\partial \mbox{\boldmath $B$}}{\partial t}+
c\mbox{\boldmath $\nabla$}\times\mbox{\boldmath $E$}=0,
\label{eq:03}
\end{equation}
\begin{equation}
\frac{\partial}{\partial t}\left(
\frac{P_{\rm g}}{\gamma_{\rm g} -1}+\frac{1}{2}\rho\mbox{\boldmath $V^2$}+
\frac{\mbox{\boldmath $B^2$}}{8\pi}\right)+
\mbox{\boldmath $\nabla$}\cdot\left[
\left(\frac{\gamma_{\rm g}}{\gamma_{\rm g}-1}P+
\frac{1}{2}\rho\mbox{\boldmath $V^2$}\right)\mbox{\boldmath $V$}+
\frac{c}{4\pi}\mbox{\boldmath $E$}\times\mbox{\boldmath $B$}\right]
+\mbox{\boldmath $V$}\cdot \left(\mbox{\boldmath $\nabla$}P_{\rm c}
-\rho\mbox{\boldmath$g$}\right)=0,
\label{eq:04}
\end{equation}
\begin{equation}
\frac{\partial}{\partial t}\left(
\frac{P_{\rm c}}{\gamma_{\rm c} -1}\right)+
\mbox{\boldmath$\nabla$}\cdot\left(
\frac{\gamma_{\rm c}}{\gamma_{\rm c}-1}P_{\rm c}\right)\mbox{\boldmath$V$}
-\mbox{\boldmath$V$}\cdot\mbox{\boldmath$\nabla$}P_{\rm c}
-\mbox{\boldmath$\nabla$}\cdot\left[
\kappa_{\|}\mbox{\boldmath $bb$}\cdot
\mbox{\boldmath$\nabla$}\left(
\frac{P_{\rm c}}{\gamma_{\rm c}-1}\right)\right]=0,
\label{eq:05}
\end{equation}
where $P_{\rm g}$ and $P_{\rm c}$ are the gas pressure and the CR pressure,
$I$ is the unit tensor, $\gamma_{\rm g}$ and $\gamma_{\rm c}$ are
the adiabatic index for the gas and the CRs, $\kappa_{\|}$ is the CR
diffusion coefficient along the magnetic field, $\mbox{\boldmath $b$}$
is the unit vector of the magnetic field, $\mbox{\boldmath $\Omega$}$
is the rotational angular frequency, and the other symbols have their
usual meanings. In this model, self-gravity is ignored. The centrifugal
force is assumed to be balanced by other forces (e.g., radial
gravitational force of the galaxy). For simplicity, we ignore
cross-field-line diffusion of the CRs. As a matter of fact, the ratio of
perpendicular- to parallel-diffusion coefficient is quite small,
say 0.02-0.04 \citep{giacalone99,ryu03}.
We normalize these equations by the physical quantities related to
the equilibrium
model described in the next subsection.
The unit of density and velocity are the density $\rho_{\rm 0}$
and sound speed $C_{\rm S0}$ at the
mid-plane of the galactic disk in the equilibrium model.
The unit of length is the scale height without magnetic field and CRs,
$H_{\rm 0}=C_{\rm S0}^2/(\gamma_{\rm g}g_z)$, and the
unit of time is the sound crossing time over one scale height,
$H_{\rm 0}/C_{\rm S0}$.
The two dimensional calculation is carried out
in the Cartesian coordinate system ($x$, $z$), where we adopted
the approximation
$\mbox{\boldmath$\hat{x}$}=\mbox{\boldmath$\hat{\varphi}$}$ and
$\mbox{\boldmath$\hat{z}$}=\mbox{\boldmath$\hat{z}$}$ in the cylindrical
coordinate system ($r$, $\varphi$, $z$) of the galactic disk,
as did \citet{mineshige93} (see Figure \ref{fig01}).
Moreover, the calculation is carried out only in the region over the
mid-plane of the galactic disk.

\subsection{Equilibrium model}
We adopted the two temperature layered disk equilibrium model
\citep{shibata89} as the initial condition,
\begin{equation}
T(z)=T_{\rm 0}+(T_{\rm halo}-T_{\rm 0})
\frac{1}{2}\left[
{\rm tanh}\left(\frac{z-z_{\rm halo}}{w_{\rm tr}}\right)+1\right],
\label{eq:06}
\end{equation}
where the disk
temperature is $T_{\rm 0}=10^4\ {\rm K}$, the halo temperature is
$T_{\rm halo}=25\times 10^4\ {\rm K}$, the height of the disk-halo
interface is $z_{\rm halo}=900\ {\rm pc}$, and the width of the transition
layer is $w_{\rm tr}=30\ {\rm pc}$. The magnetic fields are horizontal
initially.
The density, gas pressure, and CR pressure distributions are
derived from the following equation:
\begin{equation}
\frac{d}{dz}\left[P_{\rm g}+P_{\rm c}+\frac{B^2(z)}{8\pi}\right]+\rho g_z=0,
\label{eq:07}
\end{equation}
subsequently, the total gas pressure scale height at
$z=0$ (mid-plane of the galactic disk) is
$H=(1+\alpha+\beta)C_{\rm S0}^2/(\gamma_{\rm g}g_z)$, where $\alpha$,
$\beta$, and $g_z (>0)$
are the initial ratio of magnetic pressure to gas pressure,
the initial ratio of CR pressure to gas pressure, and the gravitational
acceleration, respectively.
In this paper, we only consider constant $\gamma_{\rm g}$, $\gamma_{\rm c}$,
$g_z$.
In the following simulations,
we pick $\gamma_{\rm g}=1.05$ and $\gamma_{\rm c}=4/3$,
and set $\alpha=1$ and $\beta=1$ initially.
The system is initially homogeneous in the $x$-direction.
For normalization, we take our units as follows: the unit of length is
$H_0=C_{\rm S0}^2/(\gamma_{\rm g}g_z)=50\ {\rm pc}$,
the unit of density is $\rho_0=1.6\times 10^{-24}\ {\rm g\ cm}^{-3}$,
the unit of velocity is $C_{\rm S0}=10\ {\rm km\ s}^{-1}$, and the unit of time
is $H_0/C_{\rm S0}\sim 5\ {\rm Myr}$, where the subscript 0 denotes the value
at the mid-plane of the galactic disk ($z=0$).

\section{LINEAR STABILITY ANALYSIS}

\subsection{Linearized equations}
We perform standard linear stability analysis on the set of equations
(\ref{eq:01})--(\ref{eq:05}).
Since the unperturbed state depends on $z$ only, the perturbed quantities
are assumed to have the form
$\xi=(\delta\rho,i\,\delta V_x,i\,\delta V_y,\delta V_z,\delta P_{\rm g},
\delta P_{\rm c},\delta B_x,\delta B_y,-i\,\delta B_z)$ and
\begin{equation}
\xi=\xi(z)\ \exp(\sigma t+ik_x+ik_y),
\label{eq:08}
\end{equation}
where $\sigma$ is the growth rate, $k_x$ and $k_y$ are the wave numbers
in $x$- and $y$-direction, respectively.
For simplicity, instead of linearizing the energy equation, we assume
an isothermal perturbation for the gas,
\begin{equation}
\delta P_{\rm g}=C_{\rm S}^2\delta\rho,
\label{eq:09}
\end{equation}
where $C_{\rm s}$ is the isothermal sound velocity.

Although there are nine perturbed quantities, it turns out that there are
seven algebraic relations among them, if we assume no cross-field-line diffusion
of cosmic rays.
First of all, the induction equation gives $(\delta B_x,\delta B_y,\delta B_z)$
in terms of $(\delta\rho,\delta V_x,\delta V_y,\delta V_z)$.
The x- and y-momentum equations then give $(\delta V_x,\delta V_y)$ in terms
of $(\delta\rho,\delta V_z,\delta P_{\rm t})$, where
$\delta P_{\rm t}=\delta P_{\rm g}+\delta P_{\rm c}+B_x\delta B_x/4\pi$.
The nice consequence of no cross-field-line diffusion is that $\delta P_{\rm c}$ can be
written in terms of $(\delta\rho,\delta V_z)$. (Note that the diffusion coefficient
is also perturbed, because it depends on the direction of the magnetic field.)
Thus $\rho$ can be written in terms of
$(\delta V_z,\delta P_{\rm t})$.
After some algebra, the continuity equation and the z-momentum equation become
\begin{equation}
\frac{d}{dz}{y_1 \brack y_2}=\left[\begin{array}{cc}
                               A_{11} & A_{12} \\
                               A_{21} & A_{22}
                             \end{array}\right]
{y_1 \brack y_2},
\label{eq:10}
\end{equation}
where
\begin{equation}
y_1\equiv \rho \delta V_z,
\label{eq:11}
\end{equation}
\begin{equation}
y_2\equiv\sigma\delta P_{\rm t}=\sigma\left(\delta P_{\rm g}+\delta P_{\rm c}+
\frac{B_x\delta B_x}{4\pi} \right),
\label{eq:12}
\end{equation}
\begin{equation}
K=\frac{\kappa_{\|}k_x^2}{\sigma},\ \ \Sigma^2=\sigma^2+V_{\rm A}^2k_x^2,
\ \ \Gamma=\frac{2\sigma\Omega}{\Sigma^2},
\ \ C_{\rm c}^2=\frac{\gamma_{\rm c}P_{\rm c}}{\rho},
\label{eq:13}
\end{equation}
\begin{equation}
D=1+\frac{C_{\rm c}^2}{1+K}+V_{\rm A}^2
\left[\frac{\sigma^2}{\Sigma^2}
+\frac{V_{\rm A}^2k_x^2\Gamma^2}{\Sigma^2(1+\Gamma^2)}\right],
\label{eq:14}
\end{equation}
\begin{eqnarray}
A_{11}&=&
\left[1-\frac{\sigma^2}{\Sigma^2}
-\frac{V_{\rm A}^2k_x^2}{\Sigma^2(1+\Gamma^2)}
\left(\Gamma^2+\frac{k_y}{k_x}\Gamma \right) \right]\frac{d\ln\rho}{dz} \nonumber \\
&&\frac{1}{D}\left[\frac{\sigma^2}{\Sigma^2}
+\frac{V_{\rm A}^2k_x^2}{\Sigma^2(1+\Gamma^2)}
\left(\Gamma^2+\frac{k_y}{k_x}\Gamma \right) \right] \nonumber \\
&&\times\left[g_z
+\left\{1+ \frac{C_{\rm c}^2}{1+K}
+V_{\rm A}^2\left(\frac{\sigma^2}{\Sigma^2}
+\frac{V_{\rm A}^2k_x^2\Gamma^2}{\Sigma^2(1+\Gamma^2)}\right)\right\}
\frac{d\ln\rho}{dz}
\right],
\label{eq:15}
\end{eqnarray}
\begin{eqnarray}
A_{12}&=&-\frac{k_x^2+k_y^2}{\Sigma^2(1+\Gamma^2)} \nonumber \\
&&-\frac{1}{D}\left[\frac{\sigma^2}{\Sigma^2}
+\frac{V_{\rm A}^2k_x^2}{\Sigma^2(1+\Gamma^2)}
\left(\Gamma^2+\frac{k_y}{k_x}\Gamma\right)\right]
\left[\frac{\sigma^2}{\Sigma^2}
+\frac{V_{\rm A}^2k_x^2}{\Sigma^2(1+\Gamma^2)}
\left(\Gamma^2-\frac{k_y}{k_x}\Gamma\right)\right],
\label{eq:16}
\end{eqnarray}
\begin{eqnarray}
A_{21}&=&-\Sigma^2+\frac{g_z}{D}\left[g_z
+\left\{1+\frac{C_{\rm c}^2}{1+K}
+V_{\rm A}^2\left(\frac{\sigma^2}{\Sigma^2}
+\frac{V_{\rm A}^2k_x^2\Gamma^2}{\Sigma^2(1+\Gamma^2)}\right)\right\}
\frac{d\ln\rho}{dz}
\right],
\label{eq:17}
\end{eqnarray}
\begin{equation}
A_{22}=-\frac{g_z}{D}
\left[\frac{\sigma^2}{\Sigma^2}+\frac{V_{\rm A}^2k_x^2}{\Sigma^2(1+\Gamma^2)}
\left(\Gamma^2-\frac{k_y}{k_x}\Gamma\right)\right].
\label{eq:18}
\end{equation}
We solve equation (\ref{eq:10}) to find the eigenmodes with given
boundary values.
Consequently, the problem converges to a boundary value problem, and
the growth rate of the perturbation is obtained as an eigenvalue.

\subsection{Boundary conditions}
We assume that the disk is symmetric with respect to the mid-plane ($z=0$).
Under this assumption, the sign of $\delta V_z$ inverts beyond the
mid-plane, on the other hand, the sign of
$\delta P_{\rm t}$ should not change.
Hence,
the perturbed value of $y_1$ and $y_2$ should
be antisymmetric and symmetric about $z=0$, respectively.
On the other hand, the Matrix A in equation (\ref{eq:10}) nearly
equals a constant in the region $z\gg H$ and the WKB solution is
applicable. Then, the asymptotic solutions, under the condition that
$y_1$ and $y_2$ should vanish at large $z$,
are written as follows \citep[e.g.,][]{horiuchi88}
\begin{equation}
y_1=\exp(\lambda z),
\label{eq:19}
\end{equation}
\begin{equation}
(\lambda-A_{11})y_1=A_{12}y_2,
\label{eq:20}
\end{equation}
where
\begin{equation}
\lambda=\frac{1}{2}\left(A_{11}+A_{22}-
\sqrt{(A_{11}-A_{22})^2+4A_{12}A_{21}}\right).
\label{eq:21}
\end{equation}
We solve the set of the two linear differential equations (eq.[\ref{eq:10}])
by the shooting method.
We integrate equation (\ref{eq:10}) from the outer boundary by using
the condition
given by equation (\ref{eq:19}) and (\ref{eq:20}), to the inner boundary ($z=0$)
to obtain a trial value of $\sigma$.
This $\sigma$ is regarded as an eigenvalue when the value of $y_1$
is small enough at $z=0$.

\subsection{Result of linear stability analysis}
In this analysis, we take the value of the CR diffusion coefficient
$\kappa_{\|}$, the ratio of CR pressure to gas pressure $\beta$,
and the rotational angular frequency $\Omega$ as parameters.
The value of $\kappa_{\|}$ is estimated as
$\sim 3\times 10^{28}\ {\rm cm^2\ s^{-1}}$
\citep{berezinsky90,ptuskin01,ryu03}.
In our units, $3\times 10^{28}\ {\rm cm^2\ s^{-1}}$ corresponds to $200$.
We thus take $\kappa_{\|}=200$, $\beta=1$, and $\Omega$=0 as a fiducial case,
and the ratio of magnetic pressure to gas pressure is taken as $\alpha=1$
(initially).

Figure \ref{fig02}$a$ shows the dispersion relations for different $\kappa_{\|}$
values. $\sigma$ is the growth rate and $k_x$ is the wave number
in the direction of the initial magnetic field.
We set $k_y=0$ for the linear analysis in this paper.
In this calculation, we set $\beta=1$, $\Omega=0$.
The growth rate becomes smaller
as $\kappa_{\|}$ decreases from the fiducial value, $\kappa_{\|}=200$.
This result well matches the one given by \citet{ryu03}.
The maximum growth rate $\sigma_{\rm max}$ for different $\kappa_{\|}$
occurs at roughly the same $k_x$.
Figure \ref{fig02}$b$ shows the dependence of the maximum growth rate
$\sigma_{\rm max}$ on $\kappa_{\|}$.
$\sigma_{\rm max}$ increases rapidly between $0<\kappa_{\|}<20$,
increases gradually between $20<\kappa_{\|}<80$, and finally,
becomes almost constant when $\kappa_{\|}>80$.

Figure \ref{fig03}$a$ shows the dispersion relations for different
$\beta$, where $\beta$ is the ratio of the CR pressure to the gas pressure.
In this calculation, we set $\kappa_{\|}=200$ and $\Omega=0$.
The case of $\beta=0$ is identical with the Parker instability case
without the effect of CRs.
As $\beta$ increases, the maximum growth rate $\sigma_{\rm max}$ increases,
and the wave number $k_{x{\rm max}}$, at which the growth rate becomes maximum,
increases as well. Moreover, $\sigma_{\rm max}$ and $k_{x{\rm max}}$ are
roughly related linearly. Figure \ref{fig03}$a$ also indicates that the short
wavelength perturbations become unstable as $\beta$ increases.
Figure \ref{fig03}$b$ shows the dependence of the maximum growth rate on
$\beta$. $\sigma_{\rm max}$ increases almost linearly in the range
from $\beta=0$ to $\beta=1$, and approaches a constant value as $\beta$
increases beyond $\beta=1$.

Figure \ref{fig04}$a$ shows the dispersion relations for different $\Omega$,
where $\Omega$ is the rotational angular frequency.
In this calculation, we set $\kappa_{\|}=200$.
As $\Omega$ increases, the growth rate of long wavelength perturbations
decreases rapidly, and $k_x{\rm max}$ becomes larger.
Figure \ref{fig04}$b$ shows the dependence of $\sigma_{\rm max}$ on
$\beta$ for different $\Omega$.

\section{TWO-DIMENSIONAL MHD SIMULATION}
\subsection{Numerical Procedure and Boundary Conditions}
We solve the set of two-dimensional nonlinear, time-dependent, compressible
ideal MHD equations, supplemented with the cosmic ray energy equation
(eq.[\ref{eq:01}]--[\ref{eq:05}]), in Cartesian coordinates.
We use the modified Lax-Wendroff scheme with artificial viscosity for the MHD
part, and the Biconjugate gradients stabilized (BiCGStab) method
for the diffusion part of the CR energy equation
in the same manner as described in \citet{yokoyama01}.
The MHD code using the Lax-Wendroff scheme
was originally developed by
\citet{shibata83} and has been extended by \citet{matsumoto96} and
\citet{hayashi96}.
To test the part of the code on diffusion, we did a test run on a simple CR
diffusion problem (see appendix).

We calculate only the region over the mid-plane of the galactic disk
in the $x$-$z$ plane, as shown in Figure \ref{fig01}.
The $x$-direction corresponds to
the azimuthal direction and the $z$-direction is in the direction of
the rotational axis of the galactic disk.
The partial derivative $\partial / \partial y$ is neglected.
We use the units in \S2.2 for our simulations, and the equilibrium
model in \S2.2 is used as the initial equilibrium background.
We then apply two types of perturbation, mechanical perturbations
and explosional perturbations. In the case of mechanical perturbations,
we add small velocity perturbations of the form
\begin{equation}
v_x=0.05\times C_{\rm S0}\ {\rm sin}\left[
\frac{2\pi x}{\lambda}\right]
\end{equation}
within the finite rectangular region,
$0<x<\lambda/2$ and $4H_0<z<8H_0$,
where
$\lambda=20H_0$ is the horizontal wavelength of the small velocity perturbation,
which is nearly equal to the most unstable wave length
derived from the linear analysis in the $\kappa_{\|}=200$ case.
The size of the simulation box is
$(x_{\rm max}\times z_{\rm max})=(80H_0\times 187H_0)$, the number of
grid points is $(N_x,\ N_z)=(101,\ 401)$, the grid size is $\Delta x=0.8H_0$
and $\Delta z=0.15H_0$, when $0\le z \le 25.0H_0$, and increases with $z$
otherwise.
We assume a symmetric boundary for
$x=0$ and $z=0$, and a free boundary for $x=x_{\rm max}$ and $z=z_{\rm max}$.
On the other hand, in the explosional perturbation case,
the cosmic ray energy ($\sim 10^{50}\ {\rm ergs}$) is put into a
cylindrical region which is placed at $(x,\ z)=(0,\ 6H_0)$
with volume $V_{\rm exp}=\pi r^2_{\rm exp}\, y_{\rm exp}$,
where $r_{\rm exp}=25\ {\rm pc}=0.5H_0$, $y_{\rm exp}=50\ {\rm pc}=H_0$.
The size of the simulation box is
$(x_{\rm max}\times z_{\rm max})=(90H_0\times 187H_0)$,
the number of grid points is $(N_x,\ N_z)=(301,\ 401)$,
the grid size is $\Delta x=\Delta z=0.15H_0$, when
$0\le x\le 35H_0$ and $0\le z\le 25H_0$, and increases with $x$ and $z$
otherwise.
We assume symmetric boundaries for $x=0$ and $z=0$, and a free boundary
for $x=x_{\rm max}$ and $z=z_{\rm max}$.

\subsection{Numerical Results for The Mechanical Perturbation Case}
In this subsection, we show the results for the mechanical
perturbation case.
In order to examine the effect of the cosmic ray diffusion coefficient on the
instability, we consider three different diffusion coefficients:
$\kappa_{\|}=10$, $\kappa_{\|}=40$, and $\kappa_{\|}=200$.
As can be seen from Figure \ref{fig02}, $\kappa_{\|}=10$,
$\kappa_{\|}=40$, and $\kappa_{\|}=200$ correspond to a small,
medium, and high growth rate, although the difference of the
growth rate corresponding to $\kappa_{\|}=40$ and that to $\kappa_{\|}=200$
is small.

Figure \ref{fig05}$a$-$c$
show the time evolution of the CR pressure distribution,
the magnetic field lines, and the velocity vectors for the
of $\kappa_{\|}=200$, $\kappa_{\|}=40$, and $\kappa_{\|}=10$ models.
The gray scale contour shows
the CR pressure distribution, white curves show the magnetic field lines,
and white vectors show the velocity vectors, where a white arrow at the upper
right corner shows a reference velocity vector ($=5\times C_{\rm S0}$).
The middle column shows the result of the three models
($\kappa_{\|}=200$, $\kappa_{\|}=40$, and $\kappa_{\|}=10$)
at the end of their linear growth
(i.e., $t=28$, $t=30$, $t=36$, respectively).
These times are decided
from Figure \ref{fig06}, which shows the time evolution of $V_x$ in each model.
The initial perturbations grow to form the characteristic loop like structures.
In the linear phase, the form of the magnetic loop is almost the same in each
model. In contrast, in the non-linear phase, its form can be very different.
In the $\kappa_{\|}=200$ model, the shape of the magnetic loop is
a beautiful omega shape. In the $\kappa_{\|}=40$ model, the omega
shape is distorted in the outer region and a double-loop forms in
the inner region. In the $\kappa_{\|}=10$ model, the major feature
is the double-loop.

In order to compare the results of the MHD simulation and the linear analysis,
we examine the temporal variation of $V_x$ at a particular point.
Figure \ref{fig06} shows the time evolution of $V_x$
at ($x$, $z$)=(5.6, 7.95).
The solid curves show the growth rate of $V_x$
(normalized to the sound velocity $C_{s0}$)
in each model: $\kappa_{\|}=200$, $\kappa_{\|}=40$, and $\kappa_{\|}=10$.
The broken lines, L1 for $\kappa_{\|}=200$, L2 for
$\kappa_{\|}=40$, and L3 for $\kappa_{\|}=10$, show
the growth rate obtained from the linear analysis.
The dash-dotted line shows the initial Alfv\'en speed.
The inclination obtained from the linear analysis agrees well with
that obtained from the MHD simulation.
The speed reaches the Alfv\'en speed and is saturated in the $\kappa_{\|}=200$ model.
On the other hand, the speed is saturated below the Alfv\'en speed
in the $\kappa_{\|}=40$ and $\kappa_{\|}=10$ models, and
the saturated speed decreases as $\kappa_{\|}$ decreases.

In Figure \ref{fig07}, the three panels in the first row show the CR pressure
distribution (gray scale contour), velocity vectors (white arrows),
and a magnetic field line (white curve), at the end of the linear growth phase
for the three models: (a) $\kappa_{\|}=200$ at $t=28t_0$,
(b) $\kappa_{\|}=40$ at $t=30t_0$, (c) $\kappa_{\|}=10$ at $t=36t_0$.
The arrow at the upper-right corner shows a reference velocity vector equal to
5 times a unit velocity vector. As the value of $\kappa_{\|}$
decreases, the expansion speed of magnetic loop becomes slower.
The three panels in the second row show the CR pressure values along
a magnetic field line depicted in the first row.
In the $\kappa_{\|}=200$ model, the CR pressure distribution
in the magnetic loop ($x<15$) is nearly uniform.
As $\kappa_{\|}$ decreases, the CR pressure profile develops some
structures. In addition, as $\kappa_{\|}$ decreases, the CR pressure
decreases at the top of the loop ($x\approx 0$)
but increases at the foot point of the loop.
The horizontal axis `$L$' is the distance along a magnetic field
depicted in the first row. $L=64.5$ ($\kappa_{\|}=200$),
$L=61.1$ ($\kappa_{\|}=40$), and
$L=57.8$ ($\kappa_{\|}=10$) at $x=50$.
The three panels in the third row
show the density distributions,
$\log_{10}(\rho/\rho_0)$,
along a magnetic field line depicted in the first row.
The distributions are
very similar in each model, but we can recognize the difference
of the distribution in the magnetic loop part.
In the $\kappa_{\|}=200$ model,
the density in the loop is lower than that in the others.
In the other models, the density increases in the loop as $\kappa_{\|}$ decreases.
The three panels in the fourth row show the absolute value of the velocity at the
position of the magnetic field line depicted in the first row.
The absolute value of velocity is large where the curvature of the magnetic
loop is large (except at the foot point).
Its maximum value becomes smaller as $\kappa_{\|}$ decreases.
The three panels in the fifth row show the velocity component
along the magnetic field line depicted in the first row.
When the value is positive/negative the velocity is
parallel/anti-parallel to the magnetic field. The falling speed of
matter along a magnetic field line is supersonic near the foot point
of the magnetic loop in the $\kappa_{\|}=200$ and $\kappa_{\|}=40$ models.
On the other hand, it is subsonic in the $\kappa_{\|}=10$ model.

Figure \ref{fig08}$a$-$c$ show the time evolution of
$V_z$ (velocity along the $z$-axis) sliced at $x=0$ for the
$\kappa_{\|}=200$,
$\kappa_{\|}=40$, and $\kappa_{\|}=10$ models.
In each model, $V_{\rm z}$ becomes maximum at the position of
the magnetic loop near the end of the linear growth,
$t\sim 29$ for $\kappa_{\|}=200$, $t\sim 31$ for $\kappa_{\|}=40$,
and $t\sim 36$ for $\kappa_{\|}=10$.
Subsequently, $V_{\rm z}$ becomes large in the halo region, because
the halo is pushed upward by the growing magnetic loops.
The growth speed of the magnetic loop finally falls, and the
growth is impeded similarly to the case without the effect of CRs
\citep[][chap.~17]{kato}.

\subsection{Numerical Results for The Explosional Perturbation Case}
In this subsection, we show the results for the explosional
perturbation case. In this case, we set a high CR pressure region
at ($x$, $z$)=($0$, $6$) with radius $0.5H_0$ as our initial perturbation.

Figure \ref{fig09} shows the time evolution of the CR pressure distribution
(gray scale contour), the magnetic fields (white curves), and the velocity
vectors (white arrows) for the models (a) $\kappa_{\|}=10$, and (b)
$\kappa_{\|}=80$.
It is recognized that the growth speed of the magnetic loop
in the $\kappa_{\|}=10$ model
is faster than the $\kappa_{\|}=80$ model at $t=12$, contrary to the
result for the mechanical perturbation case.
Subsequently, at $t=24$, the growth of the magnetic loop
in the $\kappa_{\|}=80$ model overtakes the $\kappa_{\|}=10$ model.
The expansion speed of the magnetic loop becomes very slow in the
$\kappa_{\|}=10$ model, while it is still very fast in the $\kappa_{\|}=80$ model.

In Figure \ref{fig10}, the two panels in the first row
show the initial CR pressure distribution (gray scale contour),
an initial magnetic field line (white line) in the models $\kappa_{\|}=10$
(the left column) and $\kappa_{\|}=80$ (the right column).
The two panels in the second row show the time evolution of the CR pressure
distribution along a magnetic field line depicted in the first row.
The field line threads through the explosion region.
The dotted line shows the initial distribution of the CR pressure, which
is highly localized near $x=0$. In the $\kappa_{\|}=10$ model,
the high CR pressure region is localized for a relatively long time,
because the diffusion speed is slow.
The initially localized high CR pressure pushes the matter in the $x$-direction
along the magnetic field lines rather effectively in the case of strong
coupling (i.e., small diffusion coefficient).
Thus the density drops rather rapidly, while the CR pressure decreases slowly
in the initial phase ($t < 2.5$).
When the magnetic loop penetrates into the low CR pressure region ($t > 2.5$),
the CR pressure inside the loop diminishes faster, partly because the magnetic tube
has expanded
and partly because the cosmic ray is carried by the downward flow of the matter.
As matter accumulates at the foot point of the magnetic loop, the magnetic tube
becomes thinner, and the CR pressure builds up, because of the small $\kappa_{\|}$.
When a significant
CR pressure gradient has been built up against the infall of matter,
the growth of the instability slows down.
This can be confirmed by the small differences of the density and
the CR pressure distributions between $t=20$ and $t=24$.
In the $\kappa_{\|}=80$ model, diffusion is more important.
The high CR pressure region disappears as cosmic rays diffuse
along the magnetic field. At $t=15$, the CR pressure is rather uniform
throughout the whole region. The CR pressure gradient between the top of
the loop and the foot point is significant only after $t=24$, and the
growth of the instability will not slow down until then.
The two panels in the third row show the time evolution of
the density distribution
along a magnetic field line depicted in the first row.
The field line threads through the explosion region.
In the $\kappa_{\|}=10$ model, matter is drained rapidly by
the CR pressure gradient and
a large drop in density occurs near the top of the magnetic loop
in a short time (at $t=2.5$). As time proceeds, the matter accumulates
at the foot point of the loop where a high density region is formed.
The draining rate of matter near the top of the loop reduces.
In the $\kappa_{\|}=80$ model, the density near the top of the
magnetic loop decreases very slowly, until about $t=15$.
After that, the draining rate accelerates, and the density near
the top of the loop becomes smaller than that of the $\kappa_{\|}=10$ model
at $t=24$. The two panels in the last row show the CR pressure
distribution, the velocity vectors, and the magnetic field line at
$t=24$.
We should point out that, in order to emphasize the CR pressure distribution near
the depicted magnetic field line, the gray scale used in Figure \ref{fig10} is
different from that used in Figure \ref{fig09}.





\section{SUMMARY AND DISCUSSION}
Using linear analysis and a time-dependent non-linear calculation, we
studied the Parker instability (or magnetic buoyancy instability)
with the effect of cosmic rays.
Several works
on the linear analysis of the Parker instability with the effect of CRs
have been published \citep[e.g.][]{hanasz97.1,ryu03}.
In \citet{hanasz97} the CR energy equation (including diffusion) was
not solved. In \citet{ryu03} the effect of rotation is not included,
and
only two cases of CR pressure were described (one was without CR $\beta=0$, and the
other was with the same unperturbed CR and gas pressures $\beta=1$).
Since \citet{ryu03} showed that the effect of
cross-field-line diffusion of CRs is negligible in the context of ISM,
we neglected the effect of cross-field-line diffusion in our
analysis for simplicity.

In the linear analysis,
the growth rate becomes larger as the CR diffusion coefficient
along the field line $\kappa_{\|}$ increases, and is saturated at large
$\kappa_{\|}$. This result is consistent with the result by
\citet{ryu03}. The growth rate also becomes larger when the
initial ratio of the CR pressure to gas pressure $\beta$ increases,
and is saturated at large $\beta$. This is consistent with \citet{ryu03},
except for some slight differences.
In our results, the maximum growth rate of the normal Parker case ($\beta=0$)
is almost half of the fiducial case ($\beta=1$), and the critical wave number,
over which the instability is stabilized, of the normal Parker case
is about $0.7$ times of the fiducial case.
In \citet{ryu03}, the maximum growth rate of the normal Parker case
is less than half of the $\beta=1$ case, and the critical wave number
of the normal Parker case is about half of the $\beta=1$ case.
The differences perhaps come from how the normalization was taken.
In fact, we succeeded in producing their results by taking the
same scale height under the same equilibrium condition. The scale height
of the disk, $H=(1+\alpha+\beta)C_{\rm s0}^2/(\gamma_{\rm g}g_z)$,
changes with the values of $\alpha$ and $\beta$, and it takes the
value $H=2C_{\rm s0}^2/(\gamma_{\rm g}g_z)$ in the normal Parker case
($\alpha=1$, $\beta=0$), and $H=3C_{\rm s0}^2/(\gamma_{\rm g}g_z)$
in the fiducial case ($\alpha=\beta=1$). We allowed the change of
the scale height because we preferred not to change the gravitational
acceleration. This is the reason why we took the unit of length as
$H_0=C_{\rm s0}^2/(\gamma_{\rm g}g_z)$.
The effect of the rotation stabilizing the Parker instability is
similar to the case without CRs.
The $k_{x{\rm max}}$ increases as the $\Omega$ increases.
Our result for the effect of rotation is
consistent with that by \citet{hanasz97.1},
except for the difference in the region of small wave number.
The growth rate with rotation is small near $k_x=0$.
It increases linearly with the wave number in \citet{hanasz97.1}.
However, in our result the growth rate increases faster than linear,
and this is also observed in the normal Parker case
\citep[see][chap.~17]{kato}.

In the MHD simulation, we showed that
the growth rate of the instability becomes smaller
when the diffusion coefficient
$\kappa_{\|}$ becomes smaller, which agrees well with the result of
the linear analysis. At the end of the linear growth,
the morphology of the magnetic loop developed from the initial
perturbation is more or less the same in the three models
($\kappa_{\|}=200$, $\kappa_{\|}=40$, and $\kappa_{\|}=10$)
studied in the mechanical perturbation case.
However, in the non-linear stage, the magnetic loop in different models
develops into different morphologies.
From the distribution of CR pressure, density, and velocity along
a magnetic field line at the end of linear growth, we found several
characteristics.
In the case of small diffusion coefficient
(e.g., $\kappa_{\|}=10$, i.e., the coupling between the CRs and the gas is strong),
the CR pressure distribution is rather non-uniform.
CRs tend to accumulate near the foot point of the magnetic loop,
and the CR pressure gradient force towards the top of the loop becomes larger.
The falling motion of matter is then impeded by the CR pressure gradient force,
and the growth rate of the Parker instability decreases.
In the case of a large diffusion coefficient
(e.g., $\kappa_{\|}=200$, $\kappa_{\|}=40$), the falling speed of matter
along a magnetic field line exceeds the speed of sound, and a shock is formed
near the foot point of the magnetic loop.
Moreover, the CR pressure distribution along a magnetic field line
in the cases of large diffusion coefficient (e.g., $\kappa_{\|}=200$)
reminds us of the profile of CR pressure in cosmic-ray-modified shocks
\citep[e.g.,][]{drury81,ko97}.
The linear growth rate in the simulations
agrees well with that in the linear analysis.
We also found that the speed along the disk is saturated at the initial
Alfv\'en speed. This result agrees with that in the normal Parker instability
\citep[i.e., without CRs,][]{matsumoto88}.
The unperturbed state has the scale height
$H=(C_{\rm s}^2+\beta C_{\rm s}^2+V_{\rm A}^2/2)/g_z$.
When the Parker instability takes place, the gas falls down along the magnetic
field lines. The CR pressure tends to distribute uniformly along a
magnetic field line (at least in the case of large diffusion coefficient),
and its contribution to the scale height disappears. Thus the scale height
along the magnetic field lines settles to $H'=C_{\rm s}^2/g_z$ at
later times. The released gravitational energy in the form of kinetic energy
per unit mass is estimated as $V_{\rm A}^2/2$. Hence, we obtain the same
results as the normal Parker case, even when the effect of CRs is included.

The explosional perturbation case has been studied by
\citet{hanasz00}. They stated that the smaller the diffusion coefficient
the larger the growth rate of the instability.
This trend is the opposite of what we found from linear analysis and
simulation in the mechanical perturbation case.
We thus computed the explosional perturbation case for a longer time.
Our result showed that the growth rate is larger in the smaller diffusion
coefficient model only in the early stage.
The growth rate becomes smaller when compare with that of the large
diffusion coefficient model in the later stage. The growth of instability is
suspended by the CR pressure gradient force interfering with the
falling motion of the matter in the small $\kappa_{\|}$ model,
while the magnetic loop can grow up to larger scales in the large
$\kappa_{\|}$ model.




\acknowledgments
Numerical computations were carried out on VPP5000 at
NAOJ. TK and CMK are supported in part by the National Science Council,
Taiwan, R.O.C., under the grants NSC-91-2112-M-008-006,
NSC-90-2112-M-008-020, NSC-91-2112-M-008-050.



\appendix

\section{TEST CALCULATION}
In this appendix we show the result of a simple CR diffusion problem to
test the diffusion part of our numerical code.
We solved the following diffusion equation
(i.e., cosmic ray energy equation [\ref{eq:04}] with
$\mbox{\boldmath $V$}=0)$,
\begin{equation}
\frac{\partial E_{\rm c}}{\partial t}
=\mbox{\boldmath$\nabla$}\cdot\left[
\kappa_{\|}\mbox{\boldmath $bb$}\cdot
\mbox{\boldmath$\nabla$}E_{\rm c}\right],
\label{eq:23}
\end{equation}
where $E_{\rm c}$ is the CR energy,
and $\kappa_{\|}$ is the diffusion coefficient along the magnetic field.
We considered a uniform magnetic field with $x$-component only, i.e.,
$\mbox{\boldmath $B$}=(B_x,0)$.
The test calculation itself is two dimensional in the $x$-$z$ plane
but the content is the same as a one dimensional calculation in the $x$-direction,
because we just considered constant $\kappa_{\|}$ and the initial condition
depended on $x$ only.
We took the same initial condition as that used in \citet{hanasz03},
\begin{equation}
E_{\rm C0}=A\ {\rm exp}\left[-\frac{x_p^2}{w_0^2}\right],
\label{eq:24}
\end{equation}
where $w_0=\sqrt{50}$ is the initial half width of the Gausian profile,
$x_p$ is the distance from the central point of calculation region,
and $A=10$ is the value at $x_p=0$.
The number of grid points used in this calculation is
($N_x$, $N_z$)=($400$, $100$), and $\kappa_{\|}=100$.
Figure \ref{fig11} shows the initial distribution and the distribution
at $t=9.4$. In the right panel ($t=9.4$), the solid curve shows the
analytical solution and the squares show the numerical solution.
The numerical solution completely matches the analytical solution.




\clearpage


\begin{figure}
\caption{Schematic picture of the simulation model and simulation box.}
\label{fig01}
\end{figure}


\begin{figure}
\caption{(a) Dispersion relation for the Parker instability with the effect
of CRs at different $\kappa_{\|}$. $\sigma$ is the growth rate
of perturbation and $k_x$ is the wave number along the direction of
the magnetic field in the unperturbed state.
(b) The dependence of the maximum growth rate
$\sigma_{\rm max}$ on $\kappa_{\|}$.}
\label{fig02}
\end{figure}


\begin{figure}
\caption{(a) Dispersion relation for the Parker instability with the
effect of CRs at different $\beta$, where $\beta$ is the initial ratio
of CR pressure to gas pressure.
(b) The dependence of the maximum growth rate $\sigma_{\rm max}$ on $\beta$.}
\label{fig03}
\end{figure}


\begin{figure}
\caption{(a) Dispersion relation for the Parker instability with the
effect of CRs at different $\Omega$, where $\Omega$ is
the rotational angular velocity of the disk.
(b)The dependence of the maximum growth rate $\sigma_{\rm max}$ on $\beta$
for different $\Omega$.}
\label{fig04}
\end{figure}


\begin{figure}
\caption{Time evolution of the CR pressure distribution, magnetic field
         lines, and velocity vectors in the models (a) $\kappa_{\|}=200$,
         (b) $\kappa_{\|}=40$, and (c) $\kappa_{\|}=10$.
         The gray scale, the white curves, and the white vectors show
         the CR pressure distribution, magnetic field lines, and velocity
         vectors. The units of length and time are 50pc and $10^6$ years,
         respectively.}
\label{fig05}
\end{figure}


\begin{figure}
\caption{Comparison of growth rate between the results of linear analysis
         and the results of MHD simulation.
         The broken lines, L1, L2, and L3 show the power law relation
         for the models $\kappa_{\|}=200$, $\kappa_{\|}=40$,
         and $\kappa_{\|}=10$. The dash-dotted line shows the initial
         Alfv\'en speed.}
\label{fig06}
\end{figure}


\begin{figure}
\caption{The CR pressure, the density, the absolute of velocity
         and the speed along a magnetic field line depicted in the first row
         at the end of linear growth in the models (a) $\kappa_{\|}=200$,
         (b) $\kappa_{\|}=40$, and (c) $\kappa_{\|}=10$.
         $L$ is the distance along the magnetic field line.}
\label{fig07}
\end{figure}


\begin{figure}
\caption{Time evolution of $V_z$ sliced at $x=0$ in the models
         (a) $\kappa_{\|}=200$, (b) $\kappa_{\|}=40$,
         and (c) $\kappa_{\|}=10$.}
\label{fig08}
\end{figure}


\begin{figure}
\caption{Time evolution of the CR pressure distribution (gray scale),
         magnetic fields (white curves), and velocity vectors (white arrows)
         in the case of explosional perturbation.}
\label{fig09}
\end{figure}


\begin{figure}
\caption{Time evolution of the CR pressure distribution, and the density
         distribution along a magnetic field initially threading the region
         where the CR energy was injected. The left column is
         $\kappa_{\|}=10$, and the right column is $\kappa_{\|}=80$.}
\label{fig10}
\end{figure}


\begin{figure}
\caption{Test result of a simple CR diffusion problem.}
\label{fig11}
\end{figure}

\end{document}